\documentclass{article}

     \usepackage[preprint]{neurips_2021}

\usepackage[utf8]{inputenc} 
\usepackage{times}
\usepackage{helvet}
\usepackage{courier}
\usepackage{hyperref}       
\usepackage{url}            
\usepackage{booktabs}       
\usepackage{graphicx} 
\usepackage{amsfonts}       
\usepackage{nicefrac}       
\usepackage{microtype}      
\usepackage{natbib}
\usepackage{amsmath}
\usepackage{amstext}
\usepackage{algorithm}
\usepackage[noend]{algpseudocode}
\usepackage{amsthm}
\usepackage{wrapfig}
\usepackage{subfig}
\usepackage{tikz}       
\usetikzlibrary{positioning}
\usepackage{caption}

\usepackage{xcolor}
\usepackage{tabularx}
\usepackage{booktabs}

\usepackage{mathtools}

\usepackage{tikz}
\usetikzlibrary{positioning,chains,fit,shapes,calc}

\makeatletter
\g@addto@macro\normalsize{%
  \setlength\abovedisplayskip{4pt}
  \setlength\belowdisplayskip{4pt}
  \setlength\abovedisplayshortskip{4pt}
  \setlength\belowdisplayshortskip{4pt}
}
\makeatother

\title{Observational Causal Inference in Novel Diseases: A Case Study of COVID-19}

\author{
Alexander Peysakhovich \\
Meta AI Research
\And
Yin Aphinyanaphongs \\
NYU Langone 
}

\begin{document}

\maketitle

\begin{abstract}
A key issue for all observational causal inference is that it relies on an unverifiable assumption - that observed characteristics are sufficient to proxy for treatment confounding. In this paper we argue that in medical cases these conditions are more likely to be met in cases where standardized treatment guidelines do not yet exist. One example of such a situation is the emergence of a novel disease. We study the case of early COVID-19 in New York City hospitals and show that observational analysis of two important thereapeutics, anti-coagulation and steroid therapy, gives results that agree with later guidelines issued via combinations of randomized trials and other evidence. We also argue that observational causal inference cannot be applied mechanically and requires domain expertise by the analyst by showing a cautionary tale of a treatment that appears extremely promising in the data, but the result is due to a quirk of hospital policy.
\end{abstract}

%
%



%
%

\maketitle

\section{Introduction}
Evaluating the effectiveness of therapies is a primary problem in medicine. The gold standard is the use of randomized trials \citep{concato2000randomized}. However, randomized trials can be expensive and time-consuming. A second form of evidence is the use of non-randomized data (e.g. data obtained during normal hospital operation) to perform observational causal inference.

The biggest issue with observational causal inference is that patient treatment assignment is not randomized - in particular that treatment assignment and outcome are correlated aka. confounded \citep{angrist2008mostly,imbens2015causal}. Thus, unlike in a randomized trial, a positive (or negative) difference between treated and control groups does not necessarily mean that the treatment causes this difference.

A large set of observational causal inference methods (e.g. propensity score based methods \citep{rosenbaum1983central}, model adjustment based methods \citep{van2011targeted}, or doubly robust methods \citep{bang2005doubly}) attempt to deal with the issue by removing the confounding using observed characteristics of the patients. In essence this creates a synthetic treatment and control group where now treatment is `as good as random.'

The key assumption behind all of these methods is that observed characteristics are sufficient to proxy for the confounding \citep{imbens2015causal}. However, this assumption is by definition unverifiable and so in general the credibility of an effect derived by observational causal inference comes down to whether the ``deconfounding'' is likely to be sufficient (or not).

In medicine some conditions have well entrenched treatment plans. In this case, we either explicitly lack the ability to match treated patients to a `similar' control. This is because when a patient that meets criteria for treatment is not treated it is for a specific reason that makes that patient `special' (perhaps for unobservable reasons) and thus not necessarily a good match for a treated patient. 

By contrast, sometimes we do not have well entrenched treatment plans. For example, in the case of novel diseases where information changes day to day, doctors have little experience with the disease, and different doctors may choose to take different actions for similar patients. In this case we argue that observational causal inference can be extremely useful as there will be considerable overlap between treated and control distributions. 

We study the case of the early COVID-19 outbreak in New York City. During this phase there were two major causes of patient degradation - thrombotic events (clotting) \citep{cantador2020incidence} and inflammation \citep{shang2020use}. We use observational causal inference to look at the effectiveness of two therapies - aggressive anti-coagulation and the use of steroids. We show that general use of therapeutic anti-coagulation (relative to prophylactic) appears not to have obvious benefits. By contrast, we show that steroid therapy does have positive effects. This is precisely the guidance that emerged over the next year of COVID-19 treatment due to randomized trials \citep{recovery2021dexamethasone, sadeghipour2021effect, remap2021therapeutic}.

There is no statistical test for whether the de-confounding assumption holds \citep{angrist2008mostly}. Thus observational causal inference requires analysts to get their hands dirty with understanding the data generating process. We show a cautionary tale - an example of a therapy, factor XA inhibitors, where patients who receive it appear to have better outcomes, even after adjusting for other observed factors. However, this positive association is unlikely to be causal - rather, patients who are deemed well enough to be discharged from the hospital were often given a factor XA inhibitor prescription and started on the treatment in their last few days.

Taken together, we argue that observational causal inference can be especially useful in the context of novel diseases where a lack of treatment protocols creates the necessary overlap in treated/control distributions. While observational results are not guaranteed to be causal without additional assumptions, knowledge of the data generating mechanism by the analyst can make these assumptions more plausible. Observational causal inference results should be used as complements to physician knowledge to help come up with treatment plans as well as generating potential hypotheses to be verified in more expensive randomized trials.   
\section{Dataset}
We use a de-identified dataset of patients admitted to a New York area hospital system for COVID19 between March and May of 2020. 
 
For all of the patients we construct a set of covariates including demographics (age, sex, race/ethnicity, smoking status), known comorbidities (prior myocardial infarction, heart failure, vascular disease, dementia, pulmonary disease, rheumatoid disease, peptic ulcer disease, diabetes, cancer, liver disease, HIV/AIDS, and the compound Charlson Score), out-patient medications (whether the patient was taking analgesics, antihistamines, antiarthritics, antifungals, antibiotics, antiparasitics, anticoagulants, antihyperglycemics, tumor necrosis inhibiting agents, antineoplastics, antiparkison drugs, antiplatelet drugs, antivirals, cardiovascular drugs, contraceptives, cough preparations, CNS drugs, autonomic drugs, diuretics, gastrointestinal drugs, hormones, immunosuppressants, muscle relaxants, pre-natal vitamins, psychotherapeuric drugs, sedatives, smoking deterrents, thyroid drugs, or vitains), as well as the first lab taken within the first 36 hours of hospital stay (we use 36 hours as we have access to lab result times rather than lab order times, see below for more description of labs used), and summary statistics (mean, min, max) of vital signs (heart rate, oxygen saturation, blood pressure, temperature, respiratory rate) from the first 24 hours of hospital stay. 

\subsection{Lab Values}
Not all patients have all lab values. To deal with this we first drop all labs with more than $20\%$ missing values. As an exception to rule we keep D-dimer (a measure of coagulation) as a covariate in our anti-coagulation analyses. It is missing for $25\%$ of patients, but we have prior knowledge that is used by clinicians in treatment assignment for anti-coagulation. 

Many of these labs are extremely right skewed, so we use a log transformation to normalize them. In addition, we see some examples of values outside of biological plausibility due to data quality issues. To deal with these we perform a winsorization at the 99th percentile for each lab.

The final list of labs that we use are: Albumin, Alkaline Phosphate, ALT, AST, Bilirubin, Blood Urea Nitrogen, C-Reactive Protein, Calcium, Chloride, Creatinine, D-Dimer, Eosonophils, Ferritin, Hematocrit, LDH, Lymphocytes, Platelet Volume, Monocytes, Neutrophils, Potassium, Protein, Prothrombin Time, Sodium, WBC.

\subsection{Outcome Measure Construction}
As our primary outcome of interest we use 21 organ support free days (21OSFD). We chose this measure after consulting with physicians working in our hospital as this measure is employed in existing randomized trials (e.g. \href{https://clinicaltrials.gov/ct2/show/record/NCT04359277}) or \citep{abdelhady2021effect}.

The 21 organ support free days measure is the number of days that a patient does not require pressors, renal replacement therapy, hi flow oxygen, or invasive mechanical ventilation. Patients that die during the course of their stay receive a $-1$ for the measure. 

We look at patients that are admitted at least 21 days before the end of our observation period, thus we do not have censoring in our data.

\section{Treatments Studied}
We study two treatments: aggressive anti-coagulation therapy and steroids. We describe the medical details of each treatment that guide some of our analysis choices below.

\subsection{Anticoagulation}
A major driver of adverse outcomes in COVID-19 patients is thrombosis \citep{bilaloglu2020thrombosis}. The standard protocol for dealing with clotting is the use of anti-coagulation (AC). AC is typically broken down into two doses - a smaller or prophylactic dose which is used to prevent clots from forming and a larger or therapeutic dose which is typically given when a clot is already detected \citep{lloyd2008anticoagulant}.

At the time this data was collected, there was major debate about whether the prevalence of thrombotic events in COVID patients should lead to a more aggressive AC strategy. Here we evaluate the use of such an aggressive strategy. Note that almost all patients at our hospital receive some form of anticoagulation therapy during their stay - $92\%$ patients in our data receive some AC with median time to first AC dose of $8$ hours, so we are not evaluating a ``no AC'' arm.

We focus on two possible strategies: aggressive early AC versus a more conservative strategy of beginning with prophylactic and moving to a larger dose later if it is needed.  We define our treatment as individuals either receive therapeutic levels (treatment) or only prophylactic (control) levels of AC for their first $72$ hours in the hospital. We remove individuals who receive more than one level of AC during this period.\footnote{The 72 hour window was chosen after consultation with physicians. Varying the window to be 24, 48, 72, or 96 hours does not change the main results. The tradeoff is that using a larger window decreases our sample size since we remove patients that receive multiple levels of AC during the window of study.}

The definition of therapeutic AC includes all intravenous Heparin, Rivaroxban, Warfarin, Dabigatran, or high dose ($\geq$60mg) Enoxaparin. Preventative/prophylatic AC includes subcutaneous heparin with $\leq$ 15,000 units per 24 hours, or low dose (less than 60mg) Enoxaparin. The vast majority of AC treatment in our sample is Heparin. 

From speaking to clinicians, we know that D-dimer was used as an indicator of possible severe clotting during this time. We see in our data that almost all patients with extremely high (>3000 ng/mL) D-dimer levels receive aggressive AC. For this reason, we restrict our analysis to patients with D-dimer levels below $3000$. 

Overall $23 \%$ of our sample receives the aggressive treatment. Figure \ref{D-dimer2treatment} below shows the observed probability of aggressive treatment by baseline D-dimer as well as distribution of D-dimer by treatment group. We see that there is substantial overlap in the `elevated' category which also makes up the majority of our patients.

\begin{figure}[!ht]
    \subfloat[D-dimer Predicts Treatment]{%
\includegraphics[scale=.45]{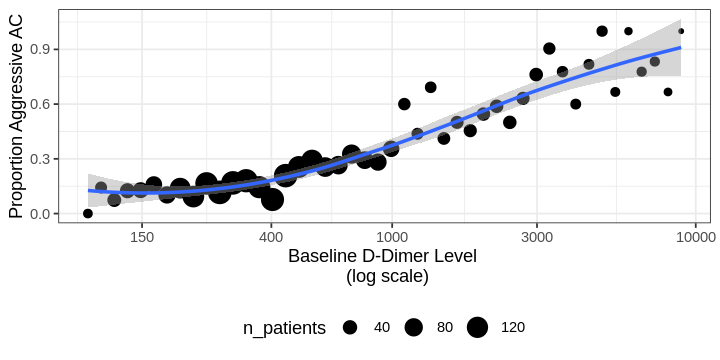}    }
    \hfill
    \subfloat[D-dimer Distribution By Treatment]{%
\includegraphics[scale=.45]{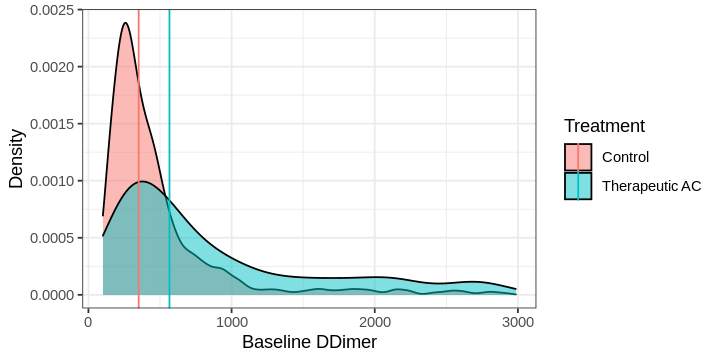}
    }
    \caption{Higher D-dimer patients were much more likely to receive aggressive AC as we would expect (panel a). We see that there is substantial overlap in D-dimer across treatments when extremely high levels ($\geq 3000$) are removed.}
    \label{D-dimer2treatment}
  \end{figure}

\subsection{Steroid Therapy}
A second major driver of adverse outcomes for COVID-19 patients is is inflammation-mediated lung injury. A treatment for inflammation is the use of glucocorticoids (steroids). In our analysis patient is deemed to be receiving steroid treatment if they receive steroids (Dexamethasone, Hydrocortisone, or Prednisone) within their first 72 hours of hospital admission.

Here our complete cases analysis includes $2282$ patients with $190$ (approximately $8\%$) of them receiving the  steroid treatment. Unlike in the AC analysis we do not have exclusion criteria based on lab results as we did not learn of any single lab that was used by all clinicians. Rather, combinations of factors led some (but not others) to treat using steroids.

The prevailing consensus at the time of the COVID19 outbreak in New York City was mixed \citep{shang2020use,russell2020clinical} with randomized trials \citep{recovery2021dexamethasone} only available much later. Though after randomized trials the use of steroids became common, in our sample, however, we see that only $\sim 8 \%$ of our patients receive steroids in their first 72 hours of hospital stay.

\section{Results}
For both treatment we look at 3 analyses: first, the overall (unadjusted) correlation between treatment and outcome. Second, we adjust for our observed covariates in a linear regression and report the coefficient on treatment. Third, we perform propensity score matching.

To do the propensity score matching we use a logistic regression on the full set of covariates to construct the propensity scores. Because most units are not treated, we matched 3 control units per 1 treated unit with a caliper of size $.05$. All analyses were done using the R package Matching \citep{sekhon2008multivariate}.

We estimate the average treatment effect on the treated (ATT) in our propensity score match. Recent work as shown that regression in observational data estimates something close to the ATT \citep{sloczynski2020interpreting} when the proportion of treated individuals is small (as in both our cases), so this makes our estimates comparable.

\subsection{Overlap Analysis}
We investigate the quality of our matching analysis by looking at overlaps in propensity scores as well as pre-and-post match balance on the covariates. Because we have a large number of covariates, we show only `important' ones in the balance plot. We select these important covariates by running two cross-validated L1 regularized regressions using the $R$ package $glmnet$ \citep{hastie2016introduction} predict the outcome and the treatment. We then take the union of the set of covariates selected by these models (note that in theory only variables which affect \textit{both} the outcome and the treatment are actually confounders so this is a very conservative selection). 

\begin{figure}[!ht]
    \subfloat[Agressive AC]{%
\includegraphics[scale=.7]{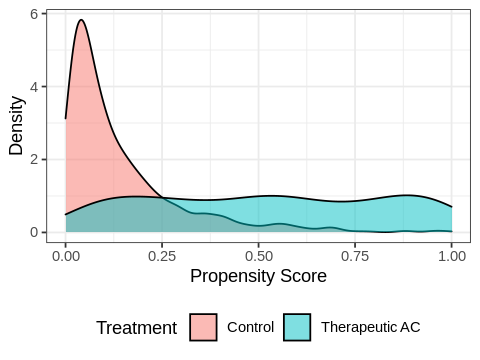}    }
    \hfill
    \subfloat[Steroid Use]{%
\includegraphics[scale=.7]{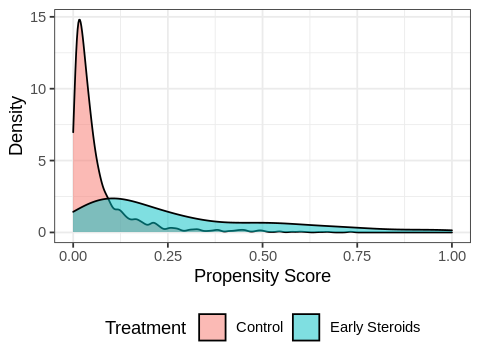}
    }
    \caption{Propensity score distributions from our dataset. We are able to match $66 \%$ of aggressive AC treated patients and $77 \%$ of steroid treated patients when using a caliper of size $.05.$}\label{prop_scores}
    \label{fig:prop}
  \end{figure}

Figure \ref{fig:prop} shows the propensity score distributions for both treatments. We see that there are some unmatchable samples (examination of our data shows that these examples appear to be very sick patients that receive aggressive treatment). However, there is a large overlap in distributions which allows for matching. We drop $84$ (approximately $33 \%$) treated patients in our matching for AC analysis and $44$ ($\sim 23 \%$) treated patients in our steroid analysis.

%
%
%

\begin{figure}[!ht]
    \subfloat[Agressive AC]{%
\includegraphics[scale=.45]{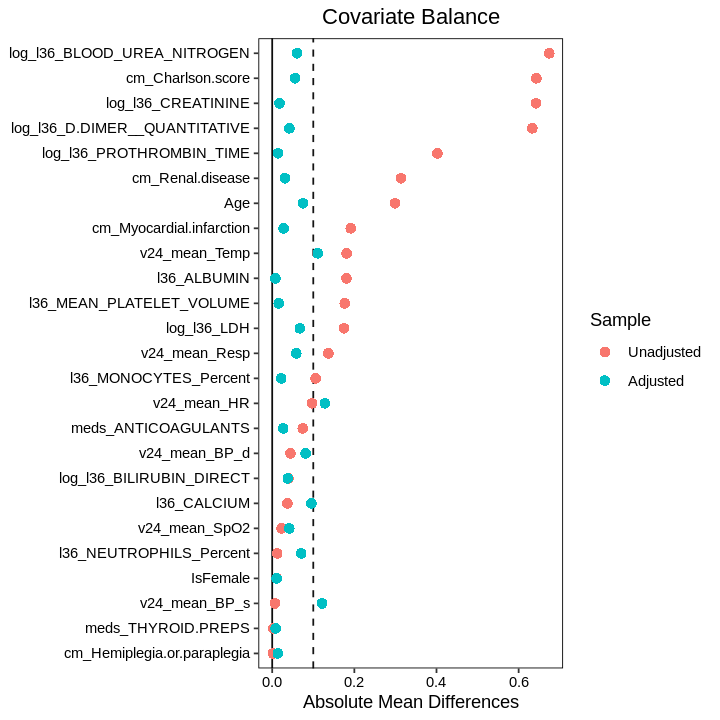}    }
    \hfill
    \subfloat[Steroid Use]{%
\includegraphics[scale=.45]{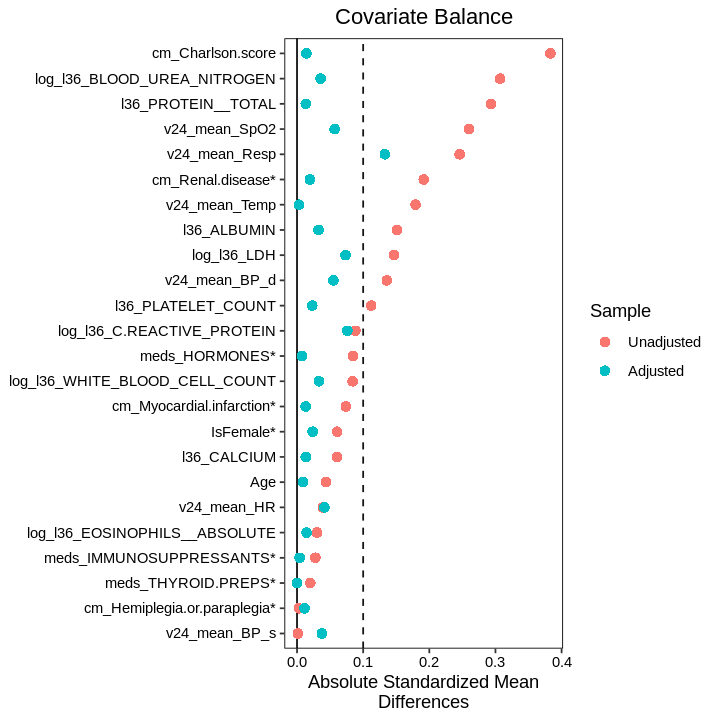}
    }
    \caption{Balance plots for our analyses. Propensity scores are trained with all labs and vital signs, to save space the balance plots only show important features as selected by Lasso linear models trained to predict treatment and outcome. Features starting with $v24$ are vital signals with BP representing blood pressure, HR representing heart rate. Features with $l36$ are lab values. Features starting with $cm$ are comorbidities. Features starting with $meds$ are outpatient medications.}
    \label{fig:bal}
  \end{figure}

Figure \ref{fig:bal} shows balance plots\footnote{We used the $R$ package cobalt (\url{https://cran.r-project.org/web/packages/cobalt/index.html}) to generate these plots.} for both treatments for the `important' variables as defined above. We see that the propensity matching produces relatively comparable distributions in terms of means in both treatments.

\subsection{Causal Effect Estimates}
\begin{figure}
\begin{center}
\includegraphics[scale=.9]{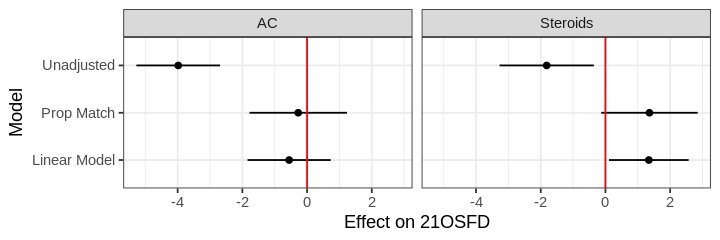}
\caption{In both AC and steroid treatments sicker patients are more likely to receive the treatment. However, after adjustment or matching we see that there is no strong association between AC treatment and outcome and a positive association between steroid treatment and outcome. Error bars indicate $95 \%$ confidence intervals.}\label{fig:results}
\end{center}
\end{figure}

Finally, we show our causal effect estimates in Figure \ref{fig:results}. We see that without balance/adjustment that there is a baseline negative association between both treatments and outcome. Thus, baseline sicker patients are more likely to receive both treatments.

However, after adjustment or matching we see that there is now a positive association between steroid treatment and outcome  (regression point estimate $1.34$ and $95\%$ CI = $[.11,.257]$ $p<.05$, matching point estimate $1.36$ and $95\%$ CI = $[-.13, 2.85]$ $p=.07$). By contrast we see little effect of more aggressive AC therapy with point estimates near $0$ (regression point estimate $-.55$ with $95\%$ CI = $[-1.82, .75]$, matching point estimate $-.27$ with $95\%$ CI = $[-1.8 .8]$).

\subsection{External Validation}
Our observational results, derived from data in the early months of the pandemic, are consistent with treatment policies which evolved over the course of the pandemic \url{www.covid19treatmentguidelines.nih.gov/therapies/antithrombotic-therapy/}) as well as by randomized trial data showing no beneficial effect of starting aggressive therapeutic level AC compared to standard care \citep{sadeghipour2021effect}. Other observational studies hint at heterogeneous effects \citep{paranjpe2020association}. We lack the statistical power to look for anything other than a main effect in our population (recall the rule of thumb that testing a single heterogeneous cut requires about $16 \times$ the data compared to testing a main effect \citep{gelman_2018}). 

By contrast, steroids (in particular, Dexamethasone) have been shown to be effective at reducing adverse events in COVID patients in randomized trials \citep{recovery2021dexamethasone} with other observational work looking for heterogeneous effects \citep{lengerich2021neutrophil}. Steroid therapy is a standard tool in practitioners treating COVID patients \url{https://www.covid19treatmentguidelines.nih.gov/management/clinical-management/hospitalized-adults--therapeutic-management/}).
\section{Conclusion and Caution}
While the prior sections may seem to suggest that observational causal inference is relatively straightforward in cases where treatment plans are not set in stone, this is quite far from the truth. Here we will argue that causal inference is not simply a statistical exercise, but one which requires knowledge about the underlying data generating process from the analyst. 

We consider another class of anti-coagulant drugs, factor XA inhibitors (FXI). These drugs are different from the mostly Heparin based AC that we considered earlier in two ways: first, they affect a different part of the clotting pathway than Heparin \citep{ansell2007factor}, second they are available in oral form while most of the AC we considered before is either subcutaenous or intravenously administered. 

We consider a naive definition of treatment where we define a patient as treated $(N_{treated}=318)$ if they received any factor XA inhibitors during their time in the hospital $(N_{control} = 1963)$.

\begin{figure}[!ht]
    \subfloat[Propensity Score Balance]{%
\includegraphics[scale=.5]{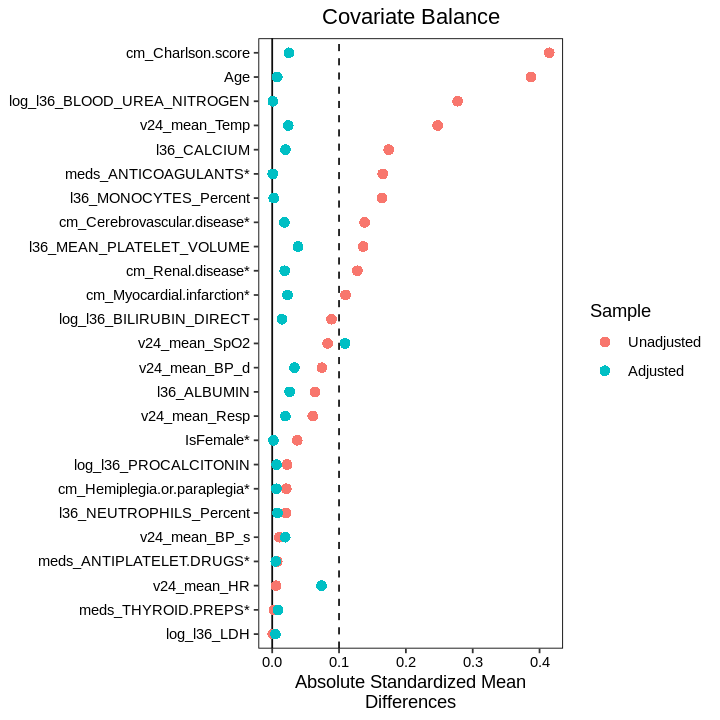}    }
    \hfill
    \subfloat[Factor XA Analysis Results]{%
\includegraphics[scale=.6]{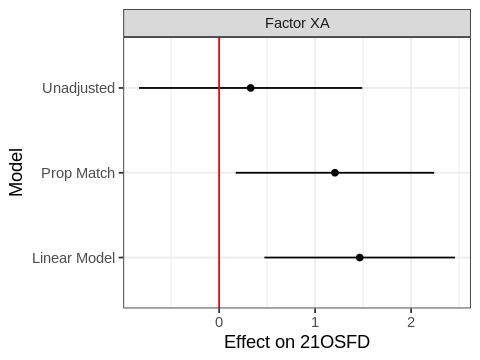}
    }
    \caption{Model results appear to show that Factor XA has a large (the size of steroids) protective affect on patients. However, this is likely an artifact of the data generating process - we know that at the time these data were collected patients who were to be discharged from the hospital were prescribed Factor XA inhibitors as out-patient treatment and started on them before the discharge occurred.}
    \label{fxa_results}
  \end{figure}

We perform the same analyses (linear model with adjusting, propensity score matching) as in the main analysis.  Figure \ref{fxa_results} shows the balance plots (as above we show only `important' variables selected via Lasso regressions) as well as the the effect estimates.

The model results appear to show that FXI has a large (the size of steroids) protective affect on patients. However, this is likely an artifact of the data generating process - we know that at the time these data were collected patients who were to be discharged from the hospital were prescribed Factor XA inhibitors as out-patient treatment and started on them before the discharge occurred. Note that a part of our erroneous conclusion comes from the way we define the treatment variable - by looking at individuals who \textit{ever} receive treatment rather than individuals who receive the treatment early in their stay (as in the AC/steroids analyses). 

We do note that there appear to be some hints in the results that something is off. First, we see that the baseline correlation between treatment and outcome is weakly positive. This means that in general we do not see the relationship that we see in steroid/AC treatments where more sick patients are more likely to get the treatment. While there are plausibly reasons for this, it does signal us to be somewhat cautious. Second, we see that even after matching we have an imbalance in the important covariates. This further suggests that very different individuals are receiving treatment and control. 

The key point remains that while observational causal inference can be a useful tool in medicine especially during times when treatment guidelines are not set in strong. In both the AC and steroid treatments we cleanly recover results that are confirmed by randomized control trials. However, gaining knowledge from observational studies requires carefully defining treatment variables and understanding the data generating process as our factor XA example shows.

\clearpage
\bibliographystyle{ACM-Reference-Format}
\bibliography{main_covidci.bbl}

\end{document}